\documentclass[aps,prl,twocolumn,superscriptaddress,longbibliography]{revtex4-1}

\usepackage{graphicx}
\usepackage{amsmath}
\usepackage{mathrsfs}
\usepackage{color}
\usepackage{xcolor}
\usepackage[colorlinks=true,linkcolor=blue,urlcolor=blue,citecolor=blue]{hyperref}
\usepackage{hyperref}
\usepackage{braket}
\usepackage{amssymb}
\usepackage{tikz}
\usepackage{bm}
\usepackage{bbm}
\usepackage{verbatim}
\usepackage{siunitx}
\usepackage{txfonts}

\newcommand{\abs}[1]{\lvert #1\rvert}

\begin{document}

\title{Exploring the Single-Particle Mobility Edge in a One-Dimensional Quasiperiodic Optical Lattice}

\author{Henrik P. L\"uschen}
\affiliation{Fakult\"at f\"ur Physik, Ludwig-Maximilians-Universit\"at M\"unchen, Schellingstr.\ 4, 80799 Munich, Germany}
\affiliation{Max-Planck-Institut f\"ur Quantenoptik, Hans-Kopfermann-Str.\ 1, 85748 Garching, Germany}

\author{Sebastian Scherg}
\affiliation{Fakult\"at f\"ur Physik, Ludwig-Maximilians-Universit\"at M\"unchen, Schellingstr.\ 4, 80799 Munich, Germany}
\affiliation{Max-Planck-Institut f\"ur Quantenoptik, Hans-Kopfermann-Str.\ 1, 85748 Garching, Germany}

\author{Thomas Kohlert}
\affiliation{Fakult\"at f\"ur Physik, Ludwig-Maximilians-Universit\"at M\"unchen, Schellingstr.\ 4, 80799 Munich, Germany}
\affiliation{Max-Planck-Institut f\"ur Quantenoptik, Hans-Kopfermann-Str.\ 1, 85748 Garching, Germany}

\author{Michael Schreiber}
\affiliation{Fakult\"at f\"ur Physik, Ludwig-Maximilians-Universit\"at M\"unchen, Schellingstr.\ 4, 80799 Munich, Germany}
\affiliation{Max-Planck-Institut f\"ur Quantenoptik, Hans-Kopfermann-Str.\ 1, 85748 Garching, Germany}

\author{Pranjal Bordia}
\affiliation{Fakult\"at f\"ur Physik, Ludwig-Maximilians-Universit\"at M\"unchen, Schellingstr.\ 4, 80799 Munich, Germany}
\affiliation{Max-Planck-Institut f\"ur Quantenoptik, Hans-Kopfermann-Str.\ 1, 85748 Garching, Germany}

\author{Xiao Li}
\affiliation{Condensed Matter Theory Center and Joint Quantum Institute, University of Maryland, College Park, Maryland 20742-4111, USA}

\author{S. Das Sarma}
\affiliation{Condensed Matter Theory Center and Joint Quantum Institute, University of Maryland, College Park, Maryland 20742-4111, USA}

\author{Immanuel Bloch}
\affiliation{Fakult\"at f\"ur Physik, Ludwig-Maximilians-Universit\"at M\"unchen, Schellingstr.\ 4, 80799 Munich, Germany}
\affiliation{Max-Planck-Institut f\"ur Quantenoptik, Hans-Kopfermann-Str.\ 1, 85748 Garching, Germany}

\date{\today}

\begin{abstract}
A single-particle mobility edge (SPME) marks a critical energy separating extended from localized states in a quantum system. In one-dimensional systems with uncorrelated disorder, a SPME cannot exist, since all single-particle states localize for arbitrarily weak disorder strengths. However, if correlations are present in the disorder potential, the localization transition can occur at a finite disorder strength and SPMEs become possible. In this work, we find experimental evidence for the existence of such a SPME in a one-dimensional quasi-periodic optical lattice. Specifically, we find a regime where extended and localized single-particle states coexist, in good agreement with theoretical simulations, which predict a SPME in this regime.

\end{abstract}

\maketitle

\paragraph{\bf{Introduction.---}}

In the presence of uncorrelated disorder, non-interacting systems can undergo Anderson localization~\cite{Anderson58}, resulting in an exponential localization of wavefunctions. In one and two dimensions, all eigenstates already localize at infinitesimal disorder strengths. In three dimensions, however, the transition occurs at a finite disorder strength~\cite{Abrahams79} and not all eigenstates need to localize at the same critical value. Instead, localized and extended states can coexist at different energies, which is the most prominent example of a so-called single-particle mobility edge (SPME)~\cite{Abrahams79,Lee85}: a critical energy separating localized from extended eigenstates. In three dimensions, this phenomenon was, among other systems (see Ref.~\cite{Lee85} for a review), observed in recent experiments with ultracold atoms~\cite{Semeghini15,Jendrzejewski12,McGehee13}, but the interpretation of the results has remained challenging~\cite{Pasek17}. While one-dimensional systems with uncorrelated disorder rigorously do not exhibit a SPME, as all states are localized for arbitrarily weak disorder strengths~\cite{Delyon85}, a related quantity called `effective mobility edge' has been identified in one-dimensional speckle potentials~\cite{Billy08}. This effective mobility edge emerges due to a finite correlation length in the speckle potential and separates exponentially from algebraically localized states~\cite{Billy08,Lugan09}, as compared to localized from extended states in systems with a true mobility edge.

For quasiperiodic potentials it is possible to construct models that do exhibit exact SPMEs even in one dimension~\cite{Sarma86,Sarma88,Thouless88,Sarma90,Biddle10,Biddle11,Ganeshan15,Johansson91,Sun15,Johansson15,Purkayastha17,Gong17,Gopalakrishnan17}, but so far their realization has remained out of reach for experiments. Recently, however, the existence of a SPME was predicted for the superposition of two optical lattices with incommensurate wavelengths~\cite{Boers07,XLi17}. For shallower lattices, the SPME is present in an intermediate phase, which separates the fully extended from the fully localized phase. At deeper lattice depths, where the nearest-neighbor tight-binding limit is approached, the intermediate phase shrinks and eventually vanishes~\cite{XLi17}. In this limit, the system maps onto the Aubry-Andr{\'e} Hamiltonian~\cite{Aubry80,Fallani07,Roati08,Modugno09,Schreiber15}, which does not display a SPME due to a self-duality in the model~\cite{Aubry80}.

\begin{figure}
	\centering
	\includegraphics[width=3.3in]{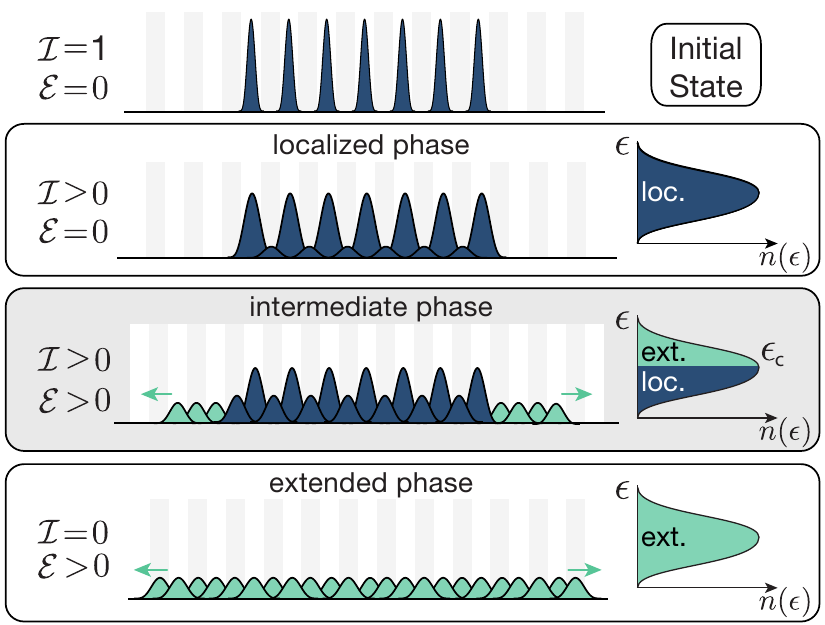}
	\caption{\textbf{Schematics of the experiment:} Schematic illustration of the initial CDW state and the states reached after time evolution in the localized, intermediate and the extended phase, respectively. The presence of localized states is marked by a persisting CDW order ($\mathcal{I} > 0$), while the presence of extended states is marked by an increase of the cloud size over time ($\mathcal{E} > 0$). In the intermediate phase, extended and localized states coexist at different energies and lead to simultaneously finite values of both $\mathcal{I} > 0$ and $\mathcal{E}>0$. As is illustrated in the diagrams of the density of states $n(\epsilon)$, they are separated by a critical energy $\epsilon_c$~\cite{XLi17}, called the mobility edge.}
	\label{Fig_schematic}
\end{figure}

In this paper, we report on the direct experimental observation of this intermediate phase in very good agreement with the theoretical predictions~\cite{XLi17}. The good agreement implies the existence of a SPME in the system, even though the critical energy itself is not directly accessible in our experiment.
We probe the intermediate phase of the bichromatic incommensurate lattice by monitoring the time evolution of an initial charge-density wave (CDW) state, as is illustrated in Fig.~\ref{Fig_schematic}. The presence of \emph{localized} states is indicated by a persisting CDW pattern for long evolution times, which is quantified via a finite density imbalance between even and odd sites $\mathcal{I} = (N_e - N_o)/(N_e+N_o)$. Here, $N_e$ ($N_o$) denote the atom number on even (odd) sites respectively. The presence of \emph{extended} states can be probed by monitoring the global size of the atom cloud $\sigma(t)$. A continuously growing expansion $\mathcal{E} \sim (\sigma(t) - \sigma(0))$ shows the presence of extended states. The intermediate phase is thus characterized by simultaneously finite values of both $\mathcal{I}$ and $\mathcal{E}$, which directly shows the coexistence of localized and extended states. Note, that the two quantities $\mathcal{I}$ and $\mathcal{E}$ are complementary in the sense that the imbalance is not sensitive to the presence of few extended states and the expansion is not sensitive to the presence of few localized states. Both quantities have been successfully utilized to study localization properties in earlier experiments~\cite{Fallani07,Roati08,Schreiber15}. Crucially, in this work, we utilize both observables simultaneously in order to detect the presence of the intermediate phase. When both indicators are finite, this implies the coexistence of both extended and localized states, which is the key ingredient of this work.

\paragraph{\bf{Experiment.---}}

In the experiment, the bichromatic optical lattice is realized via the superposition of a 
$\lambda_p \approx 532.2\,$nm `primary' lattice and a weaker incommensurate $\lambda_d \approx 738.2\,$nm `detuning' lattice at respective depths of $V_p$ and $V_d$. Deep lattices along the orthogonal directions split the system into an array of one-dimensional tubes. The system is well described by the one-dimensional Hamiltonian
\begin{equation}
\hat{H} = -\frac{\hbar^2}{2m}\frac{d^2}{dx^2} + \frac{V_p}{2} \cos{(2 k_p x)} + \frac{V_d}{2} \cos{(2 k_d x + \phi)},
\label{equ_ham}
\end{equation}
which has been studied numerically in Ref.~\cite{XLi17}.
Here, $k_i = 2 \pi/\lambda_i$ ($i=p,d$) denote the wave-vectors of the two lattices, $\phi$ the relative phase between them and $m$ the mass of the $^{40}$K atoms employed in the experiment.
We start the experiments by loading a gas of $130\times 10^3$ spin-polarized (and hence non-interacting) atoms at a temperature of $0.15 \, T_F$, into the primary and orthogonal lattices. Here, $T_F$ denotes the Fermi temperature in the dipole trap. Adding a superlattice ($\lambda_\mathrm{sup} = 1064\,$nm) to the primary lattice, the initial CDW state is created~\cite{Schreiber15}. The time-evolution is initiated by suddenly switching off the superlattice and quenching the primary and detuning lattices to their respective values. This quench results in the occupation of single-particle states throughout the entire energy spectrum.
After the time evolution, the imbalance $\mathcal{I}$ is extracted using a superlattice band-mapping technique~\cite{Trotzky12,Schreiber15}. As in previous experimental works~\cite{Schneider12,Ronzheimer13}, the size of the cloud $\sigma$ is determined from in-situ pictures and characterized by the full-width-at-half maximum (FWHM). The expansion is calculated as $\mathcal{E} = A \times (\sigma(t) - \sigma(0))$, where $A = 0.01/\mathrm{site}$ is a constant scaling factor.

We compare the experimental observables to numerical simulations. While the imbalance is directly simulated as in the experiment, the expansion is quantified via the edge density $\mathcal{D}$, which is a more direct measure of the extended states in theoretical simulations~\cite{XLi17,SOMs}. It is calculated by initially populating the eigenstates of the center third of the system before quenching to the full system. After time evolution, the edge density is calculated as $\mathcal{D} = 1- N_c/N$, where $N$ is the total particle number and $N_c$ the particle number in the originally populated center third of the system. It therefore gives the fraction of particles that leaves the originally populated center.

\paragraph{\bf{Expansion vs. edge density.---}}

\begin{figure}
	\centering
	\includegraphics[width=3.3in]{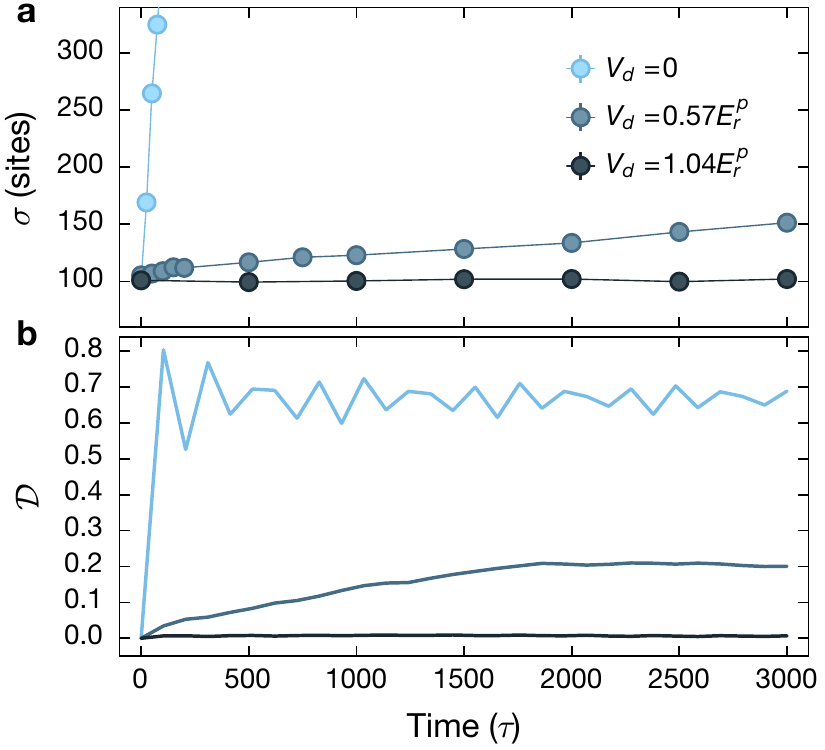}
	\caption{\textbf{Expansion versus edge density:} Time evolution of the a) experimental FWHM cloud size $\sigma$ and b) the edge density $\mathcal{D}$ obtained from numerical simulations at a primary lattice depth of $V_p = 4 E_r^p$. Data is shown in the extended ($V_d = 0$), intermediate ($V_d = 0.57 \, E_r^p$) and localized ($V_d = 1.04 \, E_r^p$) phase. Here, $E_r^p = \hbar^2 k_p^2/2m$ denotes the recoil energy of the \textit{primary} lattice and the tunneling time $\tau = \hbar/J$, where $J$ denotes the nearest neighbor tunneling rate in the primary lattice. The edge density eventually saturates due to the finite size of the simulated system.}
	\label{Fig_expansion}
\end{figure}

Fig.~\ref{Fig_expansion} compares time traces of the experimental cloud size $\sigma$ and the edge density $\mathcal{D}$ in the extended, intermediate, and localized phase.
We find that the two quantities indeed show a qualitatively similar behavior~\cite{SOMs} in describing the expansion of the system. In the extended phase, both quantities show a rapid expansion, which saturates in the numerics due to the finite size of the simulated system. In the intermediate phase, the expansion becomes dramatically slower and the numerical curve saturates to a lower value, already suggesting that not all particles are expanding. In the localized phase, neither the experiment nor the numerics shows a discernible expansion.

To enable the expansion of the cloud in the experiment, any confining (or anti-confining) potential needs to be removed. This is achieved by compensating the anti-confinement of the (blue detuned) optical lattices with the confining potential of the dipole trap to create a homogeneous potential as in Refs.~\cite{SOMs,Schneider12,Ronzheimer13}. However, the expansion dynamics in the experiment are still likely slowed down by a small, residual unevenness in the potential. This is true especially in the intermediate phase, as any unevenness becomes increasingly important in the presence of the detuning lattice~\cite{SOMs}. Still, a finite expansion remains a definite signature for the presence of extended states.

\paragraph{\bf{Results.---}}

\begin{figure*}
	\centering
	\includegraphics[width=\textwidth]{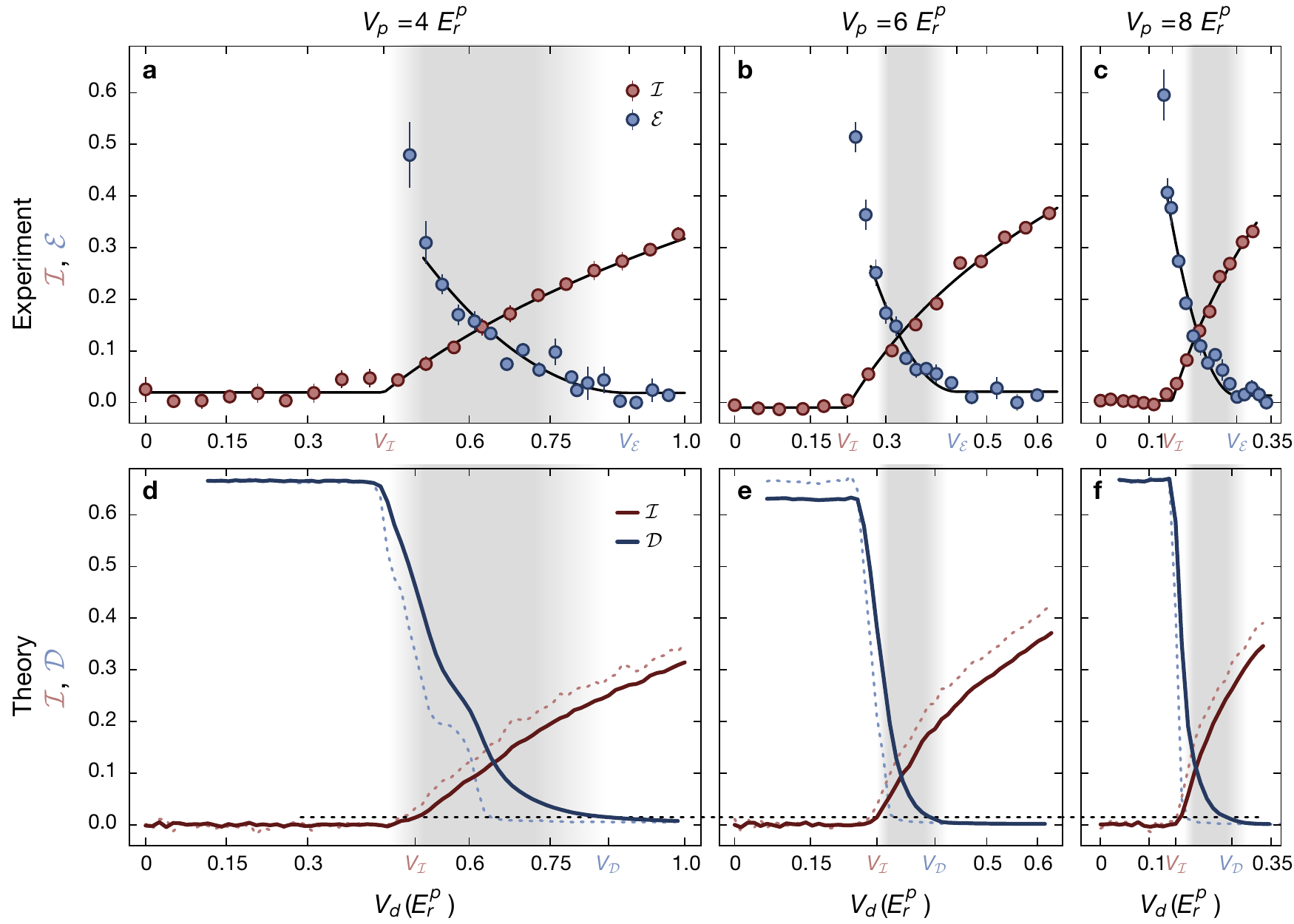}
	\caption{\textbf{Identification of the intermediate phase:} a)-c) Imbalance $\mathcal{I}$ after $200\,\tau$ and expansion $\mathcal{E}$ after $3000\,\tau$ versus detuning lattice strength $V_d$ for various depths of the primary lattice $V_p$. Experimental data is averaged over six disorder phases, the errorbars denote the standard error of the mean.  Solid lines are fitting functions to extract the critical detuning strengths for the imbalance $V_\mathcal{I}$ and the expansion $V_\mathcal{E}$.
	d)-f) Theoretically calculated imbalance $\mathcal{I}$ and edge density $\mathcal{D}$. Solid lines include the effect of averaging over many tubes with slightly different lattice depths, as is present in the experiment. Dashed lines show the result of the calculation of only the central tube (see also Ref.~\cite{XLi17}). The critical detuning strengths $V_\mathcal{I}$ ($V_\mathcal{D}$) are extracted as the points, where $\mathcal{I}$ ($\mathcal{D}$) crosses a value of $0.015$, which is marked as the black dashed horizontal line. The gray shaded region roughly marks the intermediate phase, where both the imbalances and the expansion observables are simultaneously finite and hence indicate the coexistence of extended and localized states.}
	\label{Fig_along_Vd}
\end{figure*}

\begin{figure}
	\centering
	\includegraphics[width=3.3in]{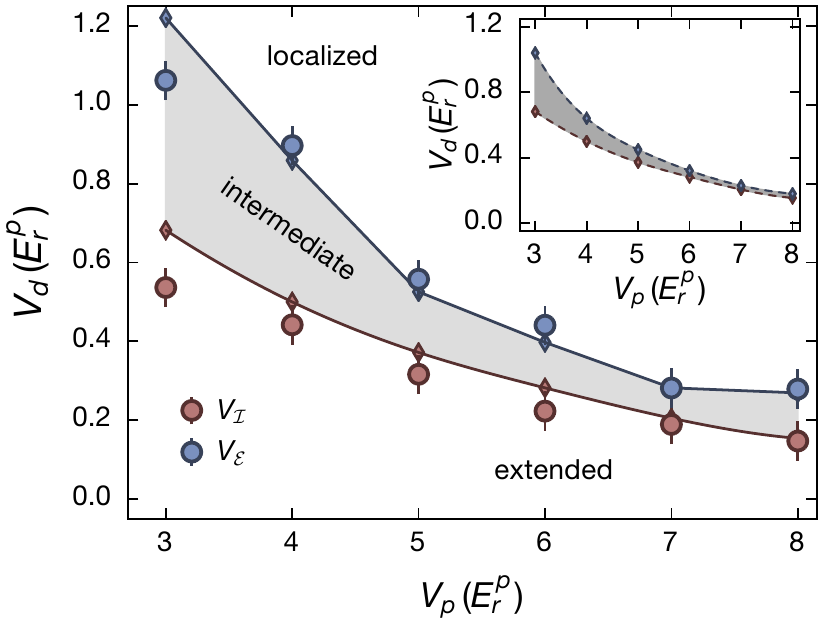}
	\caption{\textbf{Phase diagram of the incommensurate lattice model:} Boundaries of the intermediate phase (gray) as extracted from the imbalance $\mathcal{I}$ and expansion $\mathcal{E}$ from the experiment (points) and numerics (diamonds and lines) including averaging over tubes. The inset shows the numerical results for the central tube.}
	\label{Fig_phasediag}
\end{figure}

We characterize the phases of the Hamiltonian in Eq.~\eqref{equ_ham} via measurements of the imbalance and the expansion for various depths of the primary and detuning lattices $V_p$ and $V_d$ at fixed times. Due to the extremely slow expansion dynamics found in the intermediate phase (see Fig.~\ref{Fig_expansion}), we choose to extract the cloud sizes after evolution times of $3000\,\tau$. Such long evolution times are, however, not accessible for the imbalance, since it is much more susceptible to the effects of external baths, limiting its lifetime to about $T \sim 2000\,\tau$ in our case~\cite{Bordia16,Lueschen16}. Therefore, we extract the imbalance after $200\,\tau$. This is a compromise of minimizing the effects of background decays, as well as minimizing finite time errors due to slow dynamics in the intermediate phase. We note, that the imbalance is an intrinsically much faster observable than the expansion, as it does not require mass transport. In the absence of slow dynamics, it typically becomes stationary after few tunneling times already~\cite{Schreiber15}.
Even for the slow dynamics in the intermediate phase, the imbalance extracted after $200 \, \tau$ gives a reasonable estimate of its long time stationary value. We have verified this by comparing the numerically calculated imbalance after $3000\,\tau$ to the experimental value~\cite{SOMs}.

Measurements of $\mathcal{I}$ and $\mathcal{E}$ are shown in Fig.~\ref{Fig_along_Vd} a)-c).
We find that at all strengths of the primary lattice three distinct phases exist. At weak detuning lattice strengths, we always find an extended phase. It is characterized by a vanishing imbalance ($\mathcal{I} \approx 0$), which directly shows the absence of any localized states. At large detuning lattice strengths, we find a fully localized phase, which is marked by the absence of expansion ($\mathcal{E} \approx  0$). In between, a regime is found where both the imbalance and the expansion are simultaneously finite ($\mathcal{I} > 0, \mathcal{E} >0$). This directly shows the coexistence of extended and localized states, which is the defining feature of the intermediate phase, in which a SPME is present~\cite{XLi17}.

We compare our experimental results to the numerical simulations performed in Ref.~\cite{XLi17}, which are illustrated as dashed lines in Fig.~\ref{Fig_along_Vd} d)-f). We find a good agreement between the experimental and numerically simulated imbalance. However, the theoretical edge density predicts a narrower intermediate phase than the experimental expansion. We find that this difference can be explained by an averaging over many one-dimensional systems (tubes) inherently present in the experiment~\cite{SOMs}. Due to the finite extension of the beams creating the optical lattices, tubes on the outside of the system experience slightly lower lattice depths $V_p$ and $V_d$ than those in the center. The solid lines in Fig.~\ref{Fig_along_Vd} d)-f) show the numerical results including this effect. While the imbalance is only affected qualitatively, the edge density now also shows expansion up to larger detuning lattice depths as in the experiment. The stronger effect of averaging over the tubes on the edge density as compared to the imbalance is due to the first localized states emerging in the central tube with the highest lattice depths, while the last extended states vanish on the outside tubes, where the lattice depths are the lowest. The theoretical prediction of the intermediate phase including the averaging over many tubes is in very good agreement with the experimental result.

We estimate the experimental phase boundaries of the intermediate phase $V_\mathcal{I}$ and $V_\mathcal{E}$ via empirical fit functions~\cite{SOMs} to the measured imbalance and expansion, which are shown as black solid lines in Fig.~\ref{Fig_along_Vd} a)-c). Here, $V_\mathcal{I}$ denotes the lower phase boundary between the extended and the intermediate phase, which is marked by the detuning lattice depth where the imbalance first becomes finite. The upper phase boundary between the intermediate and localized phase $V_\mathcal{E}$ is at the depth of the detuning lattice where the expansion vanishes. The theoretical phase boundaries are estimated via the detuning strengths where the imbalance (or edge density) first crosses a value of $0.015$, which is just above the noise floor of the simulations. This is the same method employed in Ref.~\cite{XLi17}.
The resulting phase diagram is presented in Fig.~\ref{Fig_phasediag}. We find very good agreement between the experimental phase boundaries and the numerical calculations that include averaging over many tubes. A slight trend of the experiment to underestimate $V_\mathcal{I}$ can be attributed to finite-time effects~\cite{SOMs}. The numerical simulations not including the averaging over tubes show a smaller, but still clearly pronounced, intermediate phase (Fig.~\ref{Fig_phasediag} inset).

The intermediate phase, in which localized and extended states coexist, is most pronounced at low depths of the primary lattice $V_p$. It shrinks and shifts towards lower detuning lattice depths when the primary lattice depth is increased. In the experiment, the intermediate phase retains a small finite width even for large primary lattice depths. The comparison of numerical simulations with and without averaging over tubes shows that such a measured finite extent of the intermediate phase at e.g.\ $V_p = 8 \, E_r^p$ is almost entirely due to averaging over tubes. The intermediate phase in a single tube essentially vanishes for such primary lattice depths. Hence, in this regime, all single-particle states localize at the same critical depth of the detuning lattice with no SPME present, and the system accurately maps onto the Aubry-Andr{\'e} model~\cite{Modugno09}. The results of Fig.~\ref{Fig_phasediag} suggest that a description by the Aubry-Andr{\'e} model is approximately possible beyond primary lattice depths of $V_p > 7 \, E_r^p$, and indeed earlier experimental work on localization in the Aubry-Andr{\'e} model has been performed in this regime~\cite{Roati08,Schreiber15}.

\paragraph{\bf{Summary and Outlook.---}}
We have experimentally investigated the localization properties of a bichromatic incommensurate lattice potential over a large parameter space with non-interacting atoms. We experimentally found an intermediate phase separating the fully extended from the fully localized phase, in very good agreement with numerical simulations. In this intermediate phase, localized and extended states coexist and numerics show that a SPME is present~\cite{XLi17}. The intermediate phase vanishes in the tight-binding limit, where the lattice system maps onto the Aubry-Andr{\'e} model~\cite{Modugno09}. An experimental measurement of the critical energy separating extended from localized states would be an interesting goal for future work.

Our work presents the first experimental realization of a system with a SPME in one-dimension. Adding interactions is readily possible in our setup, opening up research prospects also in the context of many-body localization~\cite{Basko06,Iyer13,Altman15,Nandkishore15}, where couplings between localized and delocalized states via interactions might give insights into the question of the existence of a many-body mobility edge.
In fact, the possible interplay of the SPME with interaction~\cite{Li15,Modak15} remains the important open future question in this system. There are two closely related questions of fundamental importance in this problem: (1) Does many-body localization persist in the presence of a SPME as it does in the corresponding interacting Aubry-Andr{\'e} model~\cite{Iyer13,Schreiber15}? (2) Is there a many-body mobility edge in the presence of interactions? We hope to explore both questions experimentally in the future.

\begin{acknowledgments}
\paragraph{\bf{Acknowledgments.---}}
The authors thank Xiaopeng Li for discussions. 
We acknowledge financial support by the European Commission (UQUAM, AQuS) and the Nanosystems Initiative Munich (NIM). Further, this work is supported by Microsoft and the Laboratory for Physical Sciences. 
\end{acknowledgments}

\footnotetext[1]{See Supplementary Material for details}
\bibliography{SPMEBIB}

\cleardoublepage

\appendix

\setcounter{figure}{0}
\setcounter{equation}{0}

\renewcommand{\thepage}{S\arabic{page}} 
\renewcommand{\thesection}{S\arabic{section}}  
\renewcommand{\thetable}{S\arabic{table}}  
\renewcommand{\thefigure}{S\arabic{figure}}

\section{Supplementary Material}

\subsection{Comparison of different observables for the expansion}

\begin{figure*}
	\centering
	\includegraphics[width=\textwidth]{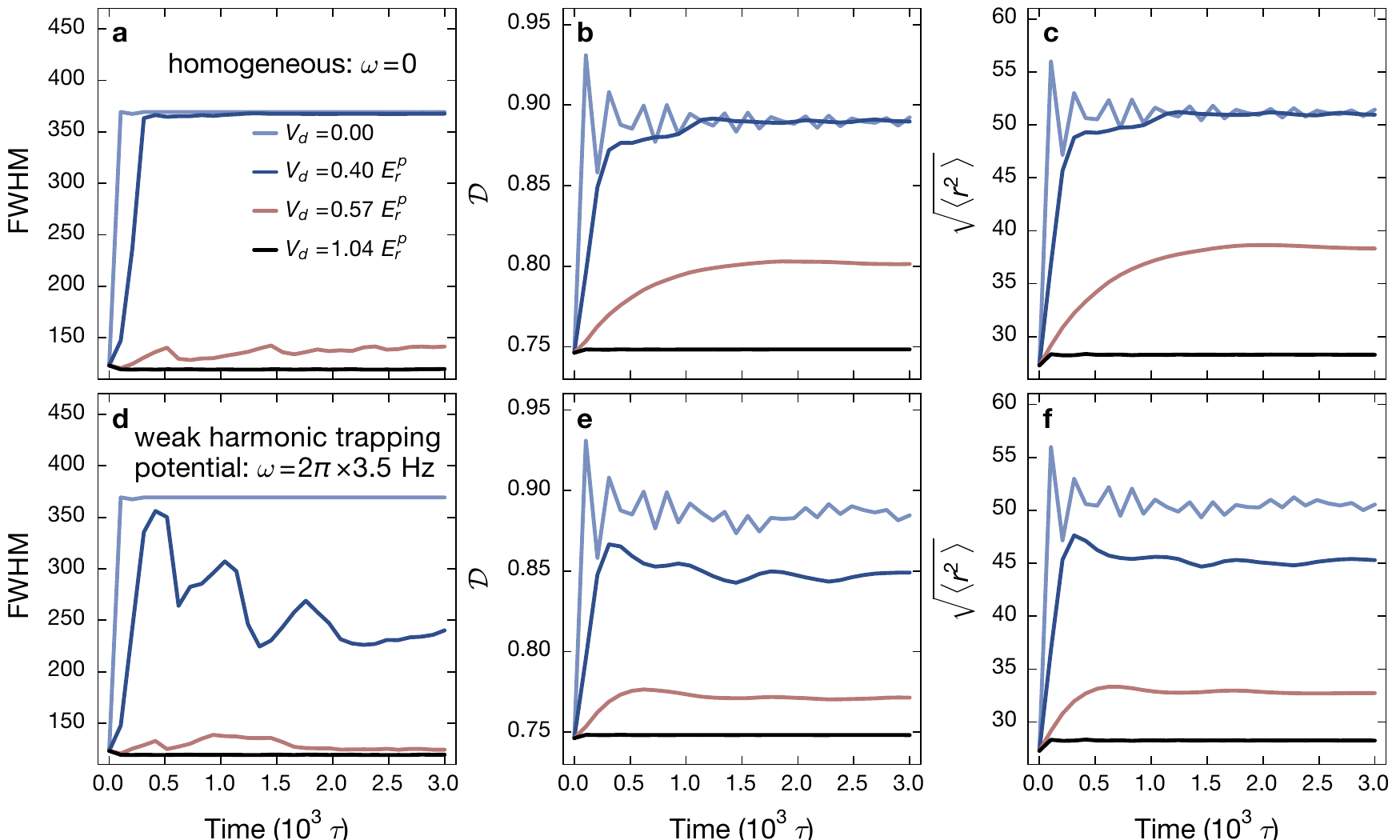}
	\caption{\textbf{Comparison of different expansion observables}: A comparison of the calculated FWHM, edge density $\mathcal{D}$, and $\sqrt{\langle r^2 \rangle }$ observable to characterize the expansion of the cloud. Traces are shown in the fully extended ($V_d = 0$ and $V_d = 0.4 \, E_r^p$), the intermediate ($V_d = 0.57 \, E_r^p$) and the fully localized ($V_d = 1.04 \, E_r^p$) phase at a primary lattice depth of $V_p = 4 \, E_r^{532\,\mathrm{nm}}$. The calculation was performed on a system of $L=369\,$sites. The initial state is a product of Wannier states with an overall Gaussian envelope with a FWHM of $\sim 123\,$ sites. a-c) Results in the absence of any confining potentials. d-f) Results in the presence of a weak trap with trapping frequency $\omega = 2 \pi \times 3.5\,$Hz. On the simulated system size, the trap results in on-site changes of the potential, which are small compared to the bandwidth. }
	\label{SOMs_Fig_expansion_observables}
\end{figure*}

Fig.~\ref{SOMs_Fig_expansion_observables}a)-c) shows a comparison of different observables characterizing the expansion in the extended, intermediate and localized phase.
The numerical data was calculated using an initial state that is very similar to the experimental initial conditions. Specifically, a product of Wannier states with an overall Gaussian envelope was chosen. Note, that this is a slightly different initial state than that used for the bulk of the computations (e.g.\ Figs.~\ref{Fig_along_Vd},~\ref{Fig_phasediag}). There, the initial state is box-like and populates only the center third of the system. Also, instead of a product of Wannier states, for the bulk of the computations a product of the eigenstates of the center third of the system was chosen. A comparison of Fig.~\ref{SOMs_Fig_expansion_observables} and Fig.~\ref{Fig_expansion}b shows that the different initial states have no qualitative effects on the expansion dynamics.

We compare three different observables characterizing the expansion: The edge density $\mathcal{D}$, the FWHM cloud size $\sigma$, and the mean square-root of the cloud size $\sqrt{\langle r^2 \rangle }$.
The edge density $\mathcal{D}$ is defined as in the main text (but calculated using the different initial state as described above). The FWHM cloud size $\sigma$ is extracted, as in the experiment, as $\sigma = \abs{i_L - i_R}$, where $i_L$ ($i_R$) are the site indices where the density first reaches half the maximum density when approaching the center from the left (right) edge of the system.
The mean square-root of the cloud size is the most often used quantity to characterize cloud sizes and calculated as $\sqrt{\langle r^2 \rangle } = \sum_i (i-i_c)^2 \langle \hat{n}_i \rangle$. Here, $i$ labels the site indices with $i_c$ being the index of the central site and $\hat{n}_i$ the density operator on site $i$.

From Fig.~\ref{SOMs_Fig_expansion_observables}, it is clearly visible that all observables related to the expansion of the cloud show the same qualitative behavior. In the extended phase, all observables quickly tend towards their respective maximum value, which is given by the system size, independent of the presence of the detuning lattice. The presence of the detuning lattice only results in a marginal slowing of the dynamics. In contrast, the expansion dynamics are significantly slower in the intermediate phase and a lower stationary value is reached. In the localized phase, no expansion is observed in any observable.

Fig.~\ref{SOMs_Fig_expansion_observables} additionally shows, that the experimentally most reliable quantity, namely the FWHM, is numerically less stable than the edge density $\mathcal{D}$ or $\sqrt{ \langle r^2 \rangle }$. This is likely due to the FWHM being highly sensitive to ripples on the density profile at the half maximum, whereas the other quantities inherently average the density of all sites. Hence, we decided to characterize the expansion in the numerical simulations via the edge density.

\subsection{Optimizing the flatness of the potential for the expansion}
Initially, the atomic cloud is strongly confined by three dipole traps traveling along the horizontal $x$- and $y$- direction ($x$ is the longitudinal direction of the tubes), as well as the vertical $z$-direction. The vertical dipole trap has a Gaussian beam waist of $\sim 150  \, \mu$m. The horizontal dipole traps have waists of $\sim 30 \, \mu$m in the vertical, but much larger waists of $\sim 300 \, \mu$m in the horizontal direction.

The optical lattices along all spatial axes have beam waists of $\sim 150 \, \mu$m (as the vertical dipole trap) and are blue detuned, providing an anti-confining potential. Note, that both a confinement and an anti-confinement can hinder the expansion of the cloud.

An approximately flat potential along the longitudinal $x$-direction is achieved by setting the dipole traps to a strength where it exactly compensates the anti-confinement of the optical lattices. As the horizontal dipole traps have different beam geometries, they cannot be used and are switched off. Hence, only the vertical dipole trap is used. 

We optimize the flatness of the potential by varying both the alignment of the vertical dipole trap, as well as its strength, to maximize the in-situ cloud size after a long evolution time (see also Refs.~\cite{Schneider12,Ronzheimer13}). Note, however, that a completely flat potential cannot be achieved, as only the harmonic contributions of the overall potential can be canceled. Additionally, small misalignments and differences in the beam shape limit the achievable flatness.

\subsection{Effect of remaining unevenness in the confining potential}

To investigate the effects of any residual unevenness on the expansion, we compare the expansion of the fully homogeneous system to the expansion in a system with a weak dipole trap using numerical simulations, as shown in Fig.~\ref{SOMs_Fig_expansion_observables}d)-f). Here, a dipole trap with a trapping frequency of $\omega = 2 \pi \times 3.5\,$Hz was used. The resulting potential at the edges of the simulated system of size $L=369\,$sites is about $V_\mathrm{dip} \approx 0.003 E_r^p$, which is approximately $1\,$\% of the bandwidth of the non-detuned system. Hence, we find that for all shown observables the expansion is not influenced in the absence of the detuning lattice. However, the presence of a weak detuning lattice can impact the expansion in the extended and the intermediate phase.

As in the experiment some leftover unevenness is unavoidable and even weak confining potentials significantly influence the dynamics, the experimental expansion measurements can deviate from those of a homogeneous system. However, the simulations presented in Fig.~\ref{SOMs_Fig_expansion_observables} also show, that the qualitative behavior, namely that the system expands in the presence of delocalized states, is not affected.

\subsection{Averaging over neighboring tubes}

While all data presented in this work is for one-dimensional (1D) systems, the experimental cloud is 3D. The 1D characteristics are achieved via two deep optical lattices along the perpendicular directions, which split the atomic cloud into a 2D array of 1D tubes. Due to the Gaussian profile of the primary and detuning lattice beams (beam waists $w \approx 150\, \mu$m), different tubes exhibit slightly different values of $V_p$ and $V_d$. To include this effect in the non-interacting simulations, we compare the beam waist to the cloud size, determined by a Gaussian fit to in-situ pictures. We obtain cloud widths of $w_y \approx 42\, \mu$m along the horizontal orthogonal direction and $w_z \approx 12 \,\mu$m along the vertical orthogonal direction. The tube averaging was carried out by performing the numerical simulations for different lattice depths and averaging the results with weights based on the number of atoms in tubes with the specified lattice depths.
\\

\subsection{Finite time effects in the imbalance}

In the experiment, the expansion of the cloud is extracted after very long evolution times of $3000\,\tau$, which is necessary due to significantly slowed down expansion dynamics in the intermediate phase. The imbalance is, however, measured after only $200 \, \tau$, as at longer times the effects of external baths become increasingly important for this observable. This potentially results in finite-time errors in the experimental imbalance.

\begin{figure}
	\centering
	\includegraphics[width=3.3in]{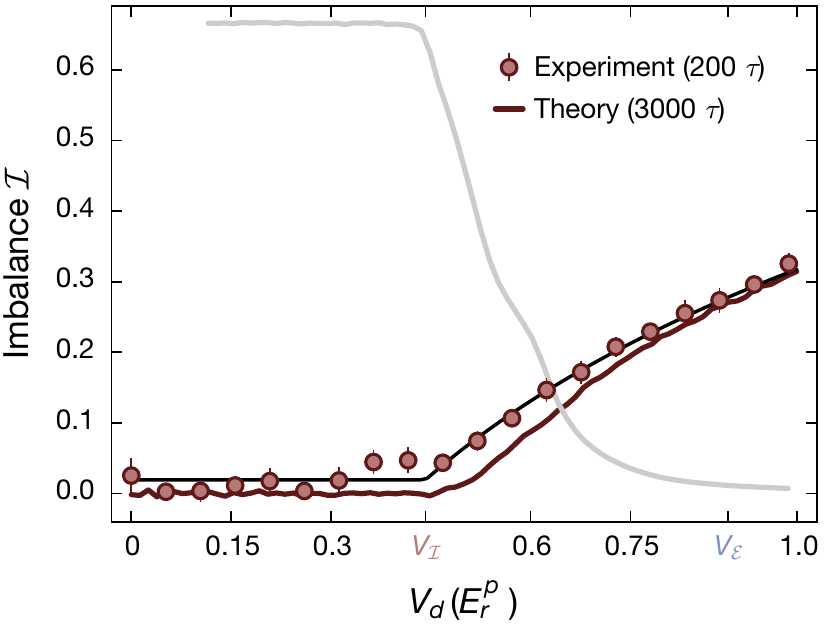}
	\caption{\textbf{Finite time error in the imbalance}: Experimentally measured and numerically calculated (including averaging over tubes) imbalance at a primary lattice depth of $V_p = 4\,E_r^p$ as in Fig.~\ref{Fig_along_Vd}a) and d). The black line illustrates the fit to the experimental data used to extract $V_\mathcal{I}$. The gray line gives the numerically calculated edge density including the averaging over tubes to illustrate the extent of the intermediate phase.}
	\label{SOMs_Fig_imbalance_finite_time}
\end{figure}

In order to estimate such a possible error, we compare the experimentally measured imbalance $\mathcal{I}$ after $200\,\tau$ to the numerically simulated imbalance (including averaging over tubes) after $3000\,\tau$ at a primary lattice depth of $V_p = 4 \, E_r^p$ in Fig.~\ref{SOMs_Fig_imbalance_finite_time}. We find, that the experimental and numerical imbalances agree very well in the fully extended and localized phase, where no slow dynamics are expected and the imbalance should become stationary after only few tunneling times~\cite{Schreiber15}. In the intermediate phase, however, we do indeed find that the experiment slightly overestimates the imbalance, most likely due to the finite time. From the fit-function used to extract the phase boundary $V_\mathcal{I}$, we can see that this finite time effect has only a small influence on the extracted value of $V_\mathcal{I}$. Still, this finite-time effect is most likely responsible for the slight underestimation of the lower boundary of the intermediate phase in Fig.~\ref{Fig_phasediag}.

\subsection{Fit-functions to extract $V_\mathcal{I}$ and $V_\mathcal{E}$}

We extract the experimental values of $V_\mathcal{I}$ and $V_\mathcal{E}$ via heuristic fit functions to the experimental data, examples of which are shown in Fig.~\ref{Fig_along_Vd}. Note, that the corresponding theoretical values are not extracted via fitting functions, but instead as the strength of the detuning lattice where $\mathcal{I}$ (or $\mathcal{D}$) crosses a threshold value of 0.015 as in Ref.~\cite{XLi17}.
\\

\noindent
The experimental imbalance is fitted as 
\begin{equation}
\mathcal{I} =
\begin{cases}
a\times \ln(V_d/V_\mathcal{I}) + o & \quad \quad V_d > V_\mathcal{I} \\
0 & \quad \quad \mathrm{else}
\end{cases}
\end{equation}
with amplitude $a$, offset $o$, and the phase boundary between the extended and the intermediate phase $V_\mathcal{I}$.
The logarithmic fit is motivated by the known behavior of the localization length in the non-interacting Aubry-Andr{\'e} model~\cite{Aubry80}. 
\\

\noindent
For the expansion, we choose the fitting function
\begin{equation}
\mathcal{E} =
\begin{cases}
b \times (V_\mathcal{E}-V_d)^2 + o & \quad \quad V_d < V_\mathcal{E} \\
0 & \quad \quad \mathrm{else}
\end{cases}
\end{equation}
with amplitude $b$, offset $o$, and the phase boundary between the intermediate and the localized phase $V_\mathcal{E}$.
The parabolic behavior was chosen as it described the data best for most values of $V_p$. For the fits to the expansion data, the fitting range had to be manually restricted, as the expansion does not follow the parabola shape below a certain $V_d$. The depth of the detuning lattice where the expansion is not well described by the parabola fit anymore is roughly at $V_\mathcal{I}$, where the imbalance becomes finite. This suggests a dramatic difference in the expansion speeds in between the delocalized and the intermediate phase, which is also observed in the theoretical calculations.

\end{document}